\documentclass[twocolumn,pre,twoside,showpacs,showkeys,preprintnumbers,floatfix]{revtex4}

\usepackage{amsmath}
\usepackage{amssymb}
\usepackage{graphicx}
\usepackage{dcolumn}
\usepackage{bm}

\def\d{\mathrm{d}}
\def\epsilon{\varepsilon}
\def\enorm#1{\left\|{#1}\right\|_2}
\def\snorm#1{\left\|{#1}\right\|_\infty}
\def\theta{\vartheta}
\def\m#1{\mathrm{#1}}
\def\Int#1#2#3{\int\limits_{#1}\!\mathrm{d}^{#2}{#3}\;}
\def\Ref#1{$(\ref{#1})$}
\def\rho{\varrho}

\def\set#1{\underline{#1}}
\def\comM{{\mathrm{C}_k\mathrm{M}}}
\def\comS{{\mathrm{C}\mathrm{A}}}
\def\vec#1{\mathbf{#1}}

\begin{document}


\title{Free Isotropic-Nematic Interfaces in Fluids of Charged Platelike Colloids}

\author{Markus Bier}
\email{bier@fluids.mpi-stuttgart.mpg.de}

\author{Ludger Harnau}

\author{S. Dietrich}

\affiliation
{
   Max-Planck-Institut f\"ur Metallforschung, \\
   Heisenbergstra\ss e 3, 70569 Stuttgart, Germany, \\
}

\affiliation
{
   Institut f\"ur Theoretische und Angewandte Physik, \\
   Universit\"at Stuttgart, Pfaffenwaldring 57, 70569 Stuttgart, Germany
}

\date{June 14, 2005}

\begin{abstract}
Bulk properties and free interfaces of mixtures of charged platelike colloids and salt are studied 
within density-functional theory. The particles are modeled by hard cuboids with their edges constrained
to be parallel to the Cartesian axes corresponding to the Zwanzig model. The charges of the particles are 
concentrated in their center. The density functional is derived by functional integration of an extension 
of the Debye-H\"uckel pair distribution function with respect to the interaction potential. For sufficiently
small macroion charges, the bulk phase diagrams exhibit one isotropic and one nematic phase separated by a 
first-order phase transition. With increasing platelet charge, the isotropic and nematic binodals are shifted 
to higher densities. The Donnan potential between the coexisting isotropic and nematic phases is inferred from 
bulk structure calculations. Non-monotonic density and nematic order parameter profiles are found at a free 
interface interpolating between the coexisting isotropic and nematic bulk phases. Moreover, electrically 
charged layers form at the free interface leading to monotonically varying electrostatic potential profiles. 
Both the widths of the free interfaces and the bulk correlation lengths are approximately given by the Debye 
length. For fixed salt density, the interfacial tension decreases upon increasing the macroion charge.
\end{abstract}



\maketitle


\section{\label{sec:intro}Introduction}

Platelike colloidal particles play a decisive, constitutive role in processes like 
agriculture, construction, oil drilling, or coating. This wide range of applicability mirrors
a very rich phase behavior of suspensions of platelets, including liquid crystalline phases, 
sol-gel transitions, and flocculation, depending on numerous material parameters, like size, 
shape, or charge of the particles, as well as on effective, solvent mediated interactions which 
can be tuned, e.g., by the choice of the solvent, salt concentration, or $p\m{H}$-value. Whereas 
bulk properties of suspensions of charged platelets have been investigated for decades \cite{Davi05}, free 
interfaces between coexisting fluid phases in such systems have not yet been studied. Here we
focus on this latter issue by proposing and studying a density-functional theory of such 
inhomogeneous multicomponent systems of charged anisotropic particles.

On the experimental side, characterization of bulk phases have been conducted for several model 
systems like natural clay \cite{Lang38,Brit69}, laponite \cite{Mour95,Gabr96,Pign97,Kroo98,
Mour98,Saun99,Levi00,Nico00,Nico01,Pori01,Pori03}, sterically stabilized gibbsite \cite{Kooi98,
Kooi01,Beek03,Beek04}, or nickel(II)hydroxide \cite{Brow98,Brow99} using methods like polarized 
light analysis \cite{Gabr96,Kooi98,Kooi01,Beek03,Beek04}, light scattering \cite{Kroo98,
Nico00,Nico01}, small-angle scattering with neutrons or x-rays \cite{Mour95,Pign97,Kroo98,Saun99,
Brow98,Brow99}, rheological measurements \cite{Mour95,Mour98,Levi00}, or NMR \cite{Pori01,Pori03}.
Since coexistence between bulk phases of charged gibbsite platelets has already been observed
experimentally \cite{Beek03,Beek04}, we expect that the spatially varying structural properties 
between them are also experimentally accessible using, e.g., scattering \cite{Diet95} or 
optical \cite{Bain98} techniques.

The theoretical description of suspensions of charged platelets is rather complicated due to
long-ranged, anisotropic interactions and many different length scales. Interfaces and surfaces 
in such systems add further difficulties induced by the partial loss of translational symmetry. 
Under these circumstances it is advisable to start with simplified models. In the presented one,
the particles are modeled as hard cuboids with pointlike charges concentrated in their center. 
Furthermore, the platelet orientations are restricted to three mutually perpendicular directions, 
which is commonly known as the Zwanzig model \cite{Zwan63}. Finally, in order to gain computational
advantages, distances between particles are not measured by the usual Euclidean norm but by the 
so-called supremum norm.

Density-functional theory \cite{Evan79} is a very effective method to investigate inhomogeneous 
fluid systems. It has recently been applied to describe suspensions of platelets with pure hard-core 
interactions near interfaces and surfaces \cite{Harn02a,Harn02b,Bier04}. Here we study platelets 
interacting via a hard-core plus a Coulomb potential; further interactions 
like dispersion forces are not considered, which corresponds to a suspension in which the indices of
refraction between solvent and solute are matched. The density functional is constructed 
by functional integration of the two-particle density with respect to the interaction potential 
\cite{Evan79}, which is the analogue to a Debye-charging process \cite{Deby23}. The two-particle 
density can be obtained, e.g., by interaction site model calculations \cite{Harn01,Harn02c} or 
expressed in terms of the potential of mean force which may be approximated by effective pair 
potentials \cite{Soga91,Rowa00,Triz02}. For reasons of computational advantages, here we choose an 
extension of the even simpler Debye-H\"uckel pair distribution function \cite{Deby23} in which the 
Debye screening factor is replaced by a spatially varying quantity. 

In view of these approximations, this theoretical model can be expected to be only qualitatively correct. 
The most subtle point is the choice of the pair distribution function, which requires further improvements 
in order to increase the quantitative reliability. However, the other parts of the formalism are expected 
to remain valid and thus provide the basis for future efforts. In this sense, the following sections 
present a generic formalism for free interfaces in fluids of charged platelike colloids, implemented exemplarily 
for the above-mentioned extended Debye-H\"uckel pair distribution function.

The text is structured as follows. In Sec.~\ref{sec:formalism} the general formalism with a 
detailed derivation of the actual density functional is presented. Section \ref{sec:bulk} 
is devoted to bulk phase diagrams of the model. The structures of free interfaces between an 
isotropic and a nematic bulk phase are calculated in Sec. \ref{sec:interface}. Section 
\ref{sec:discussion} discusses the current approach and summarizes our results.


\section{\label{sec:formalism}General formalism}

\subsection{Definitions}

\begin{figure}[thb]
   \includegraphics[scale=0.8]{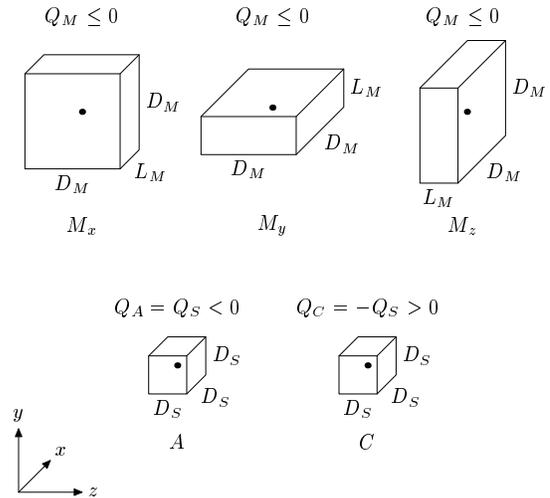}
   \caption{\label{fig:geo}Macroions $\m{M}$ are square cuboids of size $D_\m{M} \times D_\m{M} \times 
           L_\m{M}$, $D_\m{M} \not= L_\m{M}$ with charge $Q_\m{M}$, whereas anions $\m{A}$ and cations 
           $\m{C}$ are cubes of side length $D_\m{S}$ with charges $Q_\m{S}$ and $-Q_\m{S}$, respectively.
           The pointlike charges ($\bullet$) are localized in the centers of the cuboids. The macroions can adopt
           three possible orientations $\m{M_x}$, $\m{M_y}$, and $\m{M_z}$ corresponding to the 
           $L_\m{M}$-edges being parallel to the $x$-, $y$-, and $z$-axis, respectively.}
\end{figure}

We consider a ternary mixture of charged hard square cuboids with their edges required to be 
parallel to the Cartesian axes (Zwanzig model \cite{Zwan63}) dissolved in a dielectric solvent 
(e.g., water) with dielectric constant $\epsilon$. The solvent is treated as a continuum. For 
simplicity, the charges are fixed, monodisperse, and concentrated in the centers of the particles. 
The particles of the first component, representing the macroions $\m{M}$, have size 
$D_\m{M} \times D_\m{M} \times L_\m{M}$, $D_\m{M} \not= L_\m{M}$, and charge $Q_\m{M} \leq 0$. 
Within the Zwanzig approximation, macroions can take three different orientations, denoted as 
$\m{M_x}$, $\m{M_y}$, or $\m{M_z}$ corresponding to whether the $L_\m{M}$-edges are parallel to 
the $x$-, $y$-, or $z$-axis, respectively (see Fig. \ref{fig:geo}). The second component
consists of salt anions $\m{A}$ modeled as cubes of side length $D_\m{A} := D_\m{S}$ and charge 
$Q_\m{A} := Q_\m{S} < 0$ (see Fig. \ref{fig:geo}). Finally, the third component consists of salt 
cations $\m{C}$ and counterions guaranteeing overall charge neutrality. They are also described 
by cubes with the same side length $D_\m{C} := D_\m{S}$ but opposite charge $Q_\m{C} := -Q_\m{S} > 0$
(see Fig. \ref{fig:geo}).

We denote as $\rho_i(\vec{r}), i\in\{\m{M_x},\m{M_y},\m{M_z}, \m{A}, \m{C}\}$, the number density at
point $\vec{r}$ of the centers of macroions with orientation $\m{M_{x,y,z}}$, anions, and cations, 
respectively. Note that the position $\vec{r} \in V \subseteq\mathbb{R}^3$, with $V$ denoting the
system volume, is a \emph{continuous} variable in contrast to the orientation of macroions, which 
varies within a \emph{discrete} set. As an abbreviation, we introduce 
$\set{\rho}(\vec{r}) := (\rho_\m{M_x}(\vec{r}),\dots,\rho_\m{C}(\vec{r}))$. 

The system under consideration is coupled to two particle reservoirs: One supplies neutralized 
macroions (chemical formula $\displaystyle \comM, k := \frac{Q_\m{M}}{Q_\m{S}}$) and the 
other neutral salt (chemical formula $\comS$); $\mu_\comM$ and $\mu_\comS$ denote the corresponding 
chemical potentials. Upon entering the solvent, these molecules dissociate:
\begin{eqnarray}
   \m{C}_k\m{M} & \longrightarrow & k\m{C}^{Q_\m{C}} + \m{M}^{Q_\m{M}}, 
   \nonumber\\
   \m{C}\m{A}   & \longrightarrow & \m{C}^{Q_\m{C}} + \m{A}^{Q_\m{A}}.
   \label{eq:dissoc}
\end{eqnarray} 
These equilibrium chemical reactions lead to the following relations between the reservoir chemical
potentials ($\mu_\comM$ and $\mu_\comS$) and the particle chemical potentials 
($\mu_i, i\in\{\m{M_x},\m{M_y},\m{M_z}, \m{A}, \m{C}\}, \mu_\m{M_x} = \mu_\m{M_y} = \mu_\m{M_z}$): 
\begin{eqnarray}
   \mu_\comM & = & k\mu_\m{C} + \mu_\m{M_{x,y,z}} 
   \nonumber\\
   \mu_\comS & = & \mu_\m{C} + \mu_\m{A}. 
   \label{eq:murel}
\end{eqnarray}

\subsection{Density-functional theory}

The configurations of this system are characterized by the set of number density profiles $\set{\rho}$. 
The equilibrium states minimize the grand canonical density functional \cite{Evan79,Units}
\begin{equation}
   \Omega[\set{\rho}] =  
   \sum_i\Int{V}{3}{r}\rho_i(\vec{r})\left(\ln\left(\rho_i(\vec{r})\right) - 1 - \mu^*_i\right)
   + F^\m{ex}[\set{\rho}],
   \label{eq:df1}
\end{equation}
where $F^\m{ex}$ is the free energy in excess over the ideal gas contribution. Here, the reduced 
particle chemical potentials $\mu^*_i := \mu_i - \ln\left(\Lambda_i^3\right)$ with the thermal de 
Broglie wavelength $\Lambda_i$ ($\Lambda_\m{M_x} = \Lambda_\m{M_y} = \Lambda_\m{M_z}$) for particles 
of class $i$ have been introduced. With the reduced reservoir chemical potentials
\begin{eqnarray}
   \mu_\comM^* & := & \mu_\comM - \left(\ln\left(\Lambda_\m{M_{x,y,z}}^3\right) + 
                                        k \ln\left(\Lambda_\m{C}^3\right)\right)
   \nonumber\\
   \mu_\comS^* & := & \mu_\comS - \left(\ln\left(\Lambda_\m{A}^3\right) + 
                                        \ln\left(\Lambda_\m{C}^3\right)\right),
   \label{eq:resmustar}
\end{eqnarray}   
Eq. \Ref{eq:murel} takes the form
\begin{eqnarray}
   \mu^*_\comM & = & k\mu^*_\m{C} + \mu^*_\m{M_{x,y,z}} 
   \nonumber\\
   \mu^*_\comS & = & \mu^*_\m{C} + \mu^*_\m{A}. 
   \label{eq:mustarrel}
\end{eqnarray}
For given reservoir chemical potentials $\mu^*_\comM$ and $\mu^*_\comS$, the particle chemical potentials
$\mu^*_i$ are fixed by Eq. \Ref{eq:mustarrel} and the constraint of global charge neutrality:
\begin{equation}
   \Int{V}{3}{r}\sum_iQ_i\rho_i(\vec{r}) = 0.
   \label{eq:gcn}
\end{equation}

The Euler-Lagrange equations of the minimization problem read
\begin{equation}
   \frac{\delta \Omega}{\delta\rho_i(\vec{r})} = 
   \ln\left(\rho_i(\vec{r})\right) - \mu^*_i - c_i(\vec{r}) = 0
   \label{eq:ele}
\end{equation}
with the one-particle direct correlation function
\begin{equation}
   c_i(\vec{r}) := -\frac{\delta F^\m{ex}}{\delta\rho_i(\vec{r})}.
   \label{eq:1dcf}
\end{equation}

If $\set{\rho} = \set{\rho}^\m{eq}$ minimizes the density functional in Eq.~\Ref{eq:df1}, the 
grand potential $\Omega(T,V,\mu_\comM^*,\mu_\comS^*) = -p(T,V,\mu_\comM^*,\mu_\comS^*)V$ with 
the osmotic pressure $p(T,V,\mu_\comM^*,\mu_\comS^*)$ equals $\Omega[\set{\rho}^\m{eq}]$.
Phase coexistence corresponds to different states with equal values of the pressure $p$, the 
chemical potential $\mu_\comM^*$ of the reservoir of neutralized platelets, and the chemical 
potential $\mu_\comS^*$ of the salt reservoir. In particular, coexistence does \emph{not} imply 
equal values of the particle chemical potentials $\mu^*_i$, $i \in \{\m{M_x}, \m{M_y}, \m{M_z}, \m{A}, \m{C}\}$. 
Rather, coexisting bulk phases give rise to a Donnan potential maintaining different chemical potentials $\mu^*_i$.

\subsection{\label{sec:Fex}Excess free energy}

The above considerations are valid for any interaction between the particles. In this subsection, 
our choice of the model and the resulting excess free energy $F^\m{ex}$ are specified.

The interaction energy $U_{ij}(\vec{r},\vec{r'})$ of a particle of class $i$ at position 
$\vec{r}$ with a particle of class $j$ at position $\vec{r'}$ comprises a hard-core potential 
$U^\m{h}_{ij}(\vec{r},\vec{r'})$, which prevents the particles from overlapping, and a 
contribution $U^\m{c}_{ij}(\vec{r},\vec{r'})$ due to the charges: $U = U^\m{h} + U^\m{c}$. As
stated in the Introduction (Sec. \ref{sec:intro}), we do not consider dispersion forces.

The interactions between the charges are approximated as
\begin{equation}
   U^\m{c}_{ij}(\vec{r},\vec{r'}) := 
   \frac{Q_iQ_j}{\snorm{\vec{r}-\vec{r'}}},
   \label{eq:Uc}
\end{equation}
where the usual Euclidean norm $\enorm{\vec{r}} = \sqrt{x^2+y^2+z^2}$ is replaced by the supremum 
norm $\snorm{\vec{r}} = \max(|x|,|y|,|z|)$ because of computational advantages. Since these two
norms are equivalent, i.e., $\snorm{\vec{r}} \leq \enorm{\vec{r}} \leq \sqrt{3}\snorm{\vec{r}}$, 
we do not expect that the results change qualitatively due to this approximation. Furthermore, 
$\snorm{\cdot}$-spheres are cubes with their edges parallel to the Cartesian axes; therefore, the 
supremum norm is the most natural and adapted norm in the context of a Zwanzig model for cuboids.

As described in Ref. \cite{Evan79}, the exact relation
\begin{equation}
   \frac{\delta F^\m{ex}}{\delta U_{ij}(\vec{r},\vec{r'})} = 
   \frac{1}{2}\rho_i(\vec{r})\rho_j(\vec{r'})g_{ij}(\vec{r},\vec{r'})
   \label{eq:dFdU}
\end{equation}
with the pair distribution function $g$ corresponding to the pair potential $U$ can be functionally
integrated along the path $U^{(\eta)} := U^\m{h} + \eta U^\m{c}, \eta \in [0,1]$, which yields
\begin{eqnarray}
   F^\m{ex}[\set{\rho}] & = & F^\m{ex,h}[\set{\rho}] + 
                              \frac{1}{2}\sum_{ij}\Int{V}{3}{r}\Int{V}{3}{r'}
   \nonumber\\
   & & 
   \rho_i(\vec{r})\rho_j(\vec{r'})U^\m{c}_{ij}(\vec{r},\vec{r'})
   \int\limits_0^1\!\d\eta\; g^{(\eta)}_{ij}(\vec{r},\vec{r'}),
   \label{eq:Fex1}
\end{eqnarray}
where $F^\m{ex,h}$ is the excess free energy corresponding to the pure hard-core potential $U^\m{h}$, and
$g^{(\eta)}$ denotes the (inhomogeneous) pair distribution function for the pair potential 
$U^{(\eta)}$. $F^\m{ex,h}$ is chosen as the fundamental measure functional introduced by Cuesta and 
Mart\'\i nez-Rat\'on \cite{Cues97a,Cues97b}:
\begin{equation}
   F^\m{ex,h}[\set{\rho}] := \Int{V}{3}{r}\Phi\left(\set{n}(\vec{r})\right)
   \label{eq:Fexh}
\end{equation}
with the weighted densities
\begin{equation}
   n_\alpha(\vec{r}) = \sum_i\Int{V}{3}{r'} \omega_{\alpha,i}(\vec{r}-\vec{r'})\rho_i(\vec{r'})
   \label{eq:n}
\end{equation}
for $\alpha\in\{0,1x,1y,1z,2x,2y,2z,3\}$ and the excess free energy density
\begin{eqnarray}
   \Phi(\set{n}) & = &
   -n_0\ln(1-n_3) + \sum_{q\in\{x,y,z\}}\!\frac{n_{1q}n_{2q}}{1-n_3} + 
   \frac{n_{2x}n_{2y}n_{2z}}{(1-n_3)^2}.
   \nonumber\\
   & &
   \label{eq:Phi}
\end{eqnarray}

Due to Eq. \Ref{eq:Uc}, the $\eta$-integration in Eq. \Ref{eq:Fex1} may be interpreted as a
Debye charging process \cite{Deby23,McQu00}. This motivates to approximate $g^{(\eta)}$
by an expression similar to the pair distribution function of the Debye-H\"uckel theory 
\cite{Deby23,Hans86,Lee97,McQu00}:
\begin{eqnarray}
   g^{(\eta)}_{ij}(\vec{r},\vec{r'}) 
   & &
   := \exp\left(-U^\m{h}_{ij}(\vec{r},\vec{r'})\right)
   \max\left[
   0, 1 - U^\m{c}_{ij}(\vec{r},\vec{r'})
   \vphantom{\times\eta\exp\left(-\sqrt{\eta}\kappa(\vec{r},\vec{r'};[\set{\rho}])
   \snorm{\vec{r}-\vec{r'}}\right)}\right.
   \nonumber\\
   & &   
   \left.\vphantom{0, 1 - U^\m{c}_{ij}(\vec{r},\vec{r'})}
   \times\eta\exp\left(-\sqrt{\eta}\kappa(\vec{r},\vec{r'};[\set{\rho}])
   \snorm{\vec{r}-\vec{r'}}\right)
   \right].
   \label{eq:pdf}
\end{eqnarray}
This pair distribution function has non-negative values, it vanishes within the hard-core, and 
it approaches unity at infinitely large distances. Again, here the Euclidean norm $\enorm{\cdot}$ 
has been replaced by the supremum norm $\snorm{\cdot}$. The factor $\sqrt{\eta}$ in Eq. 
\Ref{eq:pdf} is introduced because in a Debye charging process all charges $Q_i$ are replaced by 
$\sqrt{\eta}Q_i$. Furthermore, the charges $Q_i$ have to be interpreted as effective charges in order
to reproduce the actual effective interactions between charged particles within Debye-H\"uckel 
theory \cite{Aubo03}. 

The screening factor $\kappa$ in Eq. \Ref{eq:pdf} is chosen as
\begin{equation}
   \kappa(\vec{r},\vec{r'};[\set{\rho}]) := 
   \frac{1}{2}\left(\tilde\kappa(\vec{r};[\set{\rho}]) + \tilde\kappa(\vec{r'};[\set{\rho}])\right)
   \label{eq:kappa}
\end{equation}
where only anions and cations contribute to the screening \cite{Warr00}:
\begin{equation}
   \tilde\kappa(\vec{r};[\set{\rho}]) := 
   \sqrt{4\pi Q_\m{S}^2(\rho_\m{A}(\vec{r}) + \rho_\m{C}(\vec{r}))}.
   \label{eq:kappatilde}
\end{equation}
Our analysis rendered that, for a spatially constant screening factor $\kappa$, the above model 
does not yield stable interfacial profiles. This led us to introduce the spatially varying 
expression in Eq. \Ref{eq:kappa}. Alternative expressions for inhomogeneous screening factors are 
known from the theory of electrolytes: In Refs. \cite{Onsa34,Lee97} non-symmetric screening 
factors are provided whereas in Ref. \cite{Groh98} $\kappa$ is calculated from the mean salt density. 
We prefer the definition in Eqs. \Ref{eq:kappa} and \Ref{eq:kappatilde} because it is symmetric and the 
screening is determined by the salt concentration \emph{at} the actually investigated positions.

The expansion of $\Omega[\set{\rho} = \set{\rho}^\m{(hom)} + \delta\set{\rho}]$ around a spatially 
homogeneous state $\set{\rho}^\m{(hom)}$ in powers of perturbations $\delta\set{\rho}$ shows that
the spatially homogeneous state $\set{\rho}^\m{(hom)}$ is unstable with respect to spatial variations 
if the macroion charge $|Q_\m{M}|$ is sufficiently large because the second order term can become negative. 
Thus, the choice for $g^{(\eta)}$ in Eq. \Ref{eq:pdf} leads to spatially \emph{inhomogeneous} bulk phases if 
the macroion charge $|Q_\m{M}|$ is larger than some threshold value. Here, we restrict ourselves to the case of 
spatially \emph{homogeneous} bulk phases, i.e., only sufficiently small macroion charges are considered.

With Eq. \Ref{eq:pdf}, the innermost integral in Eq. \Ref{eq:Fex1} can be evaluated leading to
an expression
\begin{eqnarray}
   &   & \int\limits_0^1\!\d\eta\; g^{(\eta)}_{ij}(\vec{r},\vec{r'}) 
   \label{eq:inintgrl}\\
   & = & \exp\left(-U^\m{h}_{ij}(\vec{r},\vec{r'})\right)
         \left(1 + G_{ij}(\kappa(\vec{r},\vec{r'};[\set{\rho}]), \snorm{\vec{r}-\vec{r'}})\right)
   \nonumber
\end{eqnarray}
with functions
\begin{equation}
   G_{ij}(\kappa,s) 
   :=
   -\int\limits_0^1\!\d\eta\; \min\left[1, U^\m{c}_{ij}(s)\eta\exp\left(-\sqrt{\eta}\kappa s\right)\right]
   \label{eq:Gdef}
\end{equation}
which decay for $s\longrightarrow\infty$ as
\begin{eqnarray}
   G_{ij}(\kappa,s) 
   & \simeq & 
   -\int\limits_0^1\!\d\eta\; U^\m{c}_{ij}(s)\eta\exp\left(-\sqrt{\eta}\kappa s\right)
   \nonumber\\
   & = &
   -\frac{Q_iQ_j}{s}\int\limits_0^1\!\d\eta\; \eta\exp\left(-\sqrt{\eta}\kappa s\right)
   \nonumber\\
   & = &
   -\frac{Q_iQ_j}{s}\int\limits_0^1\!\d\zeta\; 2\zeta^3\exp\left(-\zeta\kappa s\right)
   \nonumber\\
   & = &
   -\frac{2Q_iQ_j}{\kappa^4s^5}\gamma(4,\kappa s)
   \nonumber\\
   & \simeq & 
   -\frac{12Q_iQ_j}{\kappa^4 s^5},
   \label{eq:Gdecay}
\end{eqnarray}    
where $\gamma$ denotes the incomplete gamma function \cite{Grad80,Abra72}. Therefore, 
the integrand in Eq. \Ref{eq:Fex1} vanishes at small distances $\snorm{\vec{r}-\vec{r'}}$ due to
$U^\m{h}_{ij}$, whereas it decays as $\snorm{\vec{r}-\vec{r'}}^{-1}$ for 
$\snorm{\vec{r}-\vec{r'}} \longrightarrow \infty$. In order to isolate the 
$\snorm{\vec{r}-\vec{r'}}^{-1}$ asymptotics, we add and subtract unity on the right-hand side
of Eq. \Ref{eq:inintgrl}, which, after insertion into Eq. \Ref{eq:Fex1}, leads to the following 
decomposition of the excess free energy 
\begin{equation}
   F^\m{ex} = F^\m{ex,h} + F^\m{ex,c}_\m{el} + F^\m{ex,c}_\m{corr}
   \label{eq:Fex2}
\end{equation}
with the electrostatic part
\begin{eqnarray}
   F^\m{ex,c}_\m{el}[\set{\rho}] 
   & := & 
   \frac{1}{2}\sum_{ij}\Int{V}{3}{r}\Int{V}{3}{r'}\rho_i(\vec{r})\rho_j(\vec{r'})
   U^\m{c}_{ij}(\vec{r},\vec{r'})
   \nonumber\\
   & = &
   \frac{1}{2}\Int{V}{3}{r}\Int{V}{3}{r'}
   \frac{\rho^\m{Q}(\vec{r})\rho^\m{Q}(\vec{r'})}{\snorm{\vec{r}-\vec{r'}}}
   \nonumber\\
   & = &
   \frac{1}{2}\Int{V}{3}{r}\rho^\m{Q}(\vec{r})\psi(\vec{r})
   \label{eq:Fexcel}
\end{eqnarray}
and the correlation part
\begin{widetext}
\begin{eqnarray}
   F^\m{ex,c}_\m{corr}[\set{\rho}] 
   & := & 
   \frac{1}{2}\sum_{ij}\Int{V}{3}{r}\Int{V}{3}{r'}\rho_i(\vec{r})\rho_j(\vec{r'})
   U^\m{c}_{ij}(\vec{r},\vec{r'})
   \left[
   \exp\left(-U^\m{h}_{ij}(\vec{r},\vec{r'})\right)-1
   \vphantom{+ \exp\left(-U^\m{h}_{ij}(\vec{r},\vec{r'})\right)
   G_{ij}(\kappa(\vec{r},\vec{r'};[\set{\rho}]),\snorm{\vec{r}-\vec{r'}})}
   \right.
   \nonumber\\
   & & 
   \left.\vphantom{\exp\left(-U^\m{h}_{ij}(\vec{r},\vec{r'})\right)-1}
   + \exp\left(-U^\m{h}_{ij}(\vec{r},\vec{r'})\right)
   G_{ij}(\kappa(\vec{r},\vec{r'};[\set{\rho}]),\snorm{\vec{r}-\vec{r'}})
   \right].
   \label{eq:Fexccorr}
\end{eqnarray}
\end{widetext}
Here, the local charge density
\begin{equation}
   \rho^\m{Q}(\vec{r}) := \sum_iQ_i\rho_i(\vec{r})
   \label{eq:rhoc}
\end{equation}
and the electrostatic potential
\begin{equation}
   \psi(\vec{r}) := \Int{V}{3}{r'} \frac{\rho^\m{Q}(\vec{r'})}{\snorm{\vec{r}-\vec{r'}}}
   \label{eq:psi1}
\end{equation}
have been introduced. Note that, although the integrands in Eqs. \Ref{eq:Fexcel} and \Ref{eq:Fexccorr}
are undefined for $\vec{r}=\vec{r'}$, the \emph{three-dimensional} integrals exist due to the
$\snorm{\vec{r}-\vec{r'}}^{-1}$ asymptotics for $\snorm{\vec{r}-\vec{r'}}\longrightarrow 0$.
Since the integrand in Eq. \Ref{eq:Fexccorr} decays as $\snorm{\vec{r}-\vec{r'}}^{-6}$ for 
$\snorm{\vec{r}-\vec{r'}}\longrightarrow\infty$, $F^\m{ex,c}_\m{corr}$ is well-defined for all 
finite system volumes $V$ and the thermodynamic limit of the ratio $\frac{1}{V}F^\m{ex,c}_\m{corr}$ 
exists. According to the last expression in Eq. \Ref{eq:Fexcel}, the same statements are true for 
the electrostatic contribution $F^\m{ex,c}_\m{el}$ provided the electrostatic potential $\psi$ is 
well-defined. For locally charge neutral systems ($\rho^\m{Q}=0$), e.g., for bulk phases, the latter 
holds because of Eq. \Ref{eq:psi1}. In the next subsection it is shown that $\psi$ can also be 
calculated in systems with only lateral translational symmetry.

\subsection{Planar geometry}

By imposing suitable boundary conditions, we consider only systems with translational symmetry 
in the lateral $x$ and $y$ directions. Hence, in the absence of spontaneous symmetry breaking,
all densities $\rho_i$ depend at most on the $z$ coordinate. Since the thermodynamic limit of 
globally charge neutral systems of Coulomb interacting hard particles exists \cite{Lieb72}, i.e., 
the bulk free energy density depends neither on the shape nor on the boundaries of the system volume $V$, 
the following system volumes of size $2L \times 2L \times L$ are considered in the limit $L \longrightarrow \infty$:
\begin{equation}
   V(L) := A(L) \times \left[-\frac{L}{2},\frac{L}{2}\right],
   \label{eq:V}
\end{equation}
where $A(L)$ is a square in the $x$-$y$ plane of side length $2L$ with periodic boundary conditions.

The electrostatic potential in Eq. \Ref{eq:psi1} can be expressed as
\begin{equation}
   \psi(z) = \int\limits_{-\frac{L}{2}}^{\frac{L}{2}}\!\d z'\; \rho^\m{Q}(z')
             \Int{A(L)}{2}{a'}\frac{1}{\snorm{(\vec{a'},z-z')}},
   \label{eq:psi2}
\end{equation}
where $\vec{a'}$ denotes a two-dimensional vector in the $x$-$y$ plane. The inner integral in 
Eq. \Ref{eq:psi2} leads to
\begin{eqnarray}
   & &
   \Int{A(L)}{2}{a'}\frac{1}{\snorm{(\vec{a'},z-z')}} 
   \nonumber\\
   & = &
   \int\limits_0^{|x-x'|}\!\d a'\;8a'\frac{1}{|x-x'|} +
   \int\limits_{|x-x'|}^L\!\d a'\;8a'\frac{1}{a'}
   \nonumber\\
   & = &
   -4|x-x'| + 8L.
   \label{eq:surfintgrl}
\end{eqnarray}
In conjunction with the global charge neutrality constraint of Eq. \Ref{eq:gcn}, Eq. 
\Ref{eq:psi2} reduces to
\begin{equation}
   \psi(z) = -4\int\limits_{-\frac{L}{2}}^{\frac{L}{2}}\!\d z'\; \rho^\m{Q}(z')|z-z'|.
   \label{eq:psi3}
\end{equation}
By differentiating twice, one finds that $\psi$ fulfills the Poisson equation:
$\psi'' = -8\rho^\m{Q}$. Furthermore, by making use of global charge neutrality (Eq. 
\Ref{eq:gcn}), one finds
\begin{eqnarray}
   \psi\left(-\frac{L}{2}\right) 
   & = &
   -4\int\limits_{-\frac{L}{2}}^{\frac{L}{2}}\!\d z'\; \rho^\m{Q}(z')\left(z'+\frac{L}{2}\right)
   \nonumber\\
   & = &
   -4\int\limits_{-\frac{L}{2}}^{\frac{L}{2}}\!\d z'\; \rho^\m{Q}(z')z'
   \nonumber\\
   & = &
   4\int\limits_{-\frac{L}{2}}^{\frac{L}{2}}\!\d z'\; \rho^\m{Q}(z')\left(\frac{L}{2}-z'\right)
   \nonumber\\
   & = &
   -\psi\left(\frac{L}{2}\right).
   \label{eq:sympsi}
\end{eqnarray}

Thus, the density functional in Eq. \Ref{eq:df1} takes the final form
\begin{eqnarray}
   \Omega[\set{\rho}] 
   & = & 
   4L^2\sum_i\int\limits_{-\frac{L}{2}}^{\frac{L}{2}}\!\d z\; \rho_i(z)
   \left(
   \ln\left(\rho_i(z)\right) - 1 - \mu^*_i 
   \vphantom{+ \frac{1}{2}Q_i\psi(z)}
   \right. 
   \nonumber\\
   & &
   \left.
   \vphantom{\ln\left(\rho_i(z)\right) - 1 - \mu_i}
   + \frac{1}{2}Q_i\psi(z)
   \right) 
   + F^\m{ex,h}[\set{\rho}] + F^\m{ex,c}_\m{corr}[\set{\rho}]
   \label{eq:df3}
\end{eqnarray}
which has to be minimized under the constraint of global charge neutrality (Eq. \Ref{eq:gcn}).
This is carried out by numerically solving the Euler-Lagrange equations (Eq. \Ref{eq:ele}) 
with a Picard-iteration scheme on a one-dimensional grid. 

In the isotropic and nematic bulk fluid, the densities $\set{\rho}$ are spatially constant. 
In this case, the Euler-Lagrange equations Eq. \Ref{eq:ele} comprise 
five coupled equations:
\begin{equation}
   \ln(\rho_i^\m{(bulk)}) - \mu_i^{*\m{(bulk)}} - c_i^\m{h,(bulk)} - c_{\m{corr},i}^\m{c,(bulk)} 
   = 0,
   \label{eq:eleb}
\end{equation}
where the chemical potentials $\mu_i^{*\m{(bulk)}}$ fulfill Eq. \Ref{eq:mustarrel} and the
local charge neutrality condition ($\sum_i Q_i\rho_i^\m{(bulk)} = 0$). The electrostatic 
contribution $-c_{\m{el},i}^\m{c,(bulk)} = Q_i\psi^\m{(bulk)}$ is absent in Eq. \Ref{eq:eleb} because 
the electrostatic potential $\psi$ vanishes in locally charge neutral systems (see Eq. \Ref{eq:psi1}).

For determining the number density profiles at free interfaces between coexisting bulk phases $\mathcal{B}_1$
and $\mathcal{B}_2$, the Euler-Lagrange equations
\begin{equation}
   \ln\left(\rho_i(z)\right) - \mu^*_i + Q_i\psi(z) - c_i^\m{h}(z) - c_{\m{corr},i}^\m{c}(z) 
   = 0
   \label{eq:elefi}
\end{equation}
are to be solved with the boundary conditions
\begin{equation}
   \rho_i\left(z=-\frac{L}{2}\right) = \rho_i^{(\mathcal{B}_1)}, \qquad
   \rho_i\left(z=\frac{L}{2}\right) = \rho_i^{(\mathcal{B}_2)}.
   \label{eq:bc}
\end{equation}

In order that for $z=-\frac{L}{2}$ and $z=\frac{L}{2}$ Eq. \Ref{eq:elefi} reduces to Eq. 
\Ref{eq:eleb} for $\mathcal{B}_1$ and $\mathcal{B}_2$, respectively, one has the requirements
\begin{eqnarray}
   -\mu^*_i + Q_i\psi\left(-\frac{L}{2}\right) & = & -\mu_i^{*(\mathcal{B}_1)},
   \nonumber\\
   -\mu^*_i + Q_i\psi\left(\frac{L}{2}\right)  & = & -\mu_i^{*(\mathcal{B}_2)}.
   \label{eq:req}
\end{eqnarray}
Using Eq. \Ref{eq:sympsi}, one readily concludes
\begin{equation}
   \mu^*_i = \frac{1}{2}\left(\mu_i^{*(\mathcal{B}_1)} + \mu_i^{*(\mathcal{B}_2)}\right)
   \label{eq:nuii}
\end{equation}
and
\begin{equation}
   \psi_D := \psi\left(\frac{L}{2}\right)-\psi\left(-\frac{L}{2}\right) 
           = \frac{1}{Q_i}\left(\mu_i^{*(\mathcal{B}_1)} - \mu_i^{*(\mathcal{B}_2)}\right).
   \label{eq:psiD}
\end{equation}
$\psi_D$ is known as \emph{Donnan potential} \cite{Adam73} between the two bulk phases 
$\mathcal{B}_1$ and $\mathcal{B}_2$. It maintains a density gradient of the mobile particles
at the interface between two coexisting bulk phases. Its definition given above is unique, 
i.e., the rightmost expression is in fact independent of $i$ due to Eq. \Ref{eq:mustarrel}, e.g.,
\begin{eqnarray}
   \frac{\mu_\m{A}^{*(\mathcal{B}_1)} - \mu_\m{A}^{*(\mathcal{B}_2)}}{Q_\m{A}}
   & = &
   \frac{\left(\mu_\comS^* - \mu_\m{C}^{*(\mathcal{B}_1)}\right) - 
         \left(\mu_\comS^* - \mu_\m{C}^{*(\mathcal{B}_2)}\right)}{-Q_\m{C}}
   \nonumber\\
   & = &
   \frac{\mu_\m{C}^{*(\mathcal{B}_1)} - \mu_\m{C}^{*(\mathcal{B}_2)}}{Q_\m{C}}.
   \label{eq:egpsiD}
\end{eqnarray}


\section{\label{sec:bulk}Bulk fluid}

As a first step in the investigation of the density functional developed in the last 
section, bulk phase diagrams are determined for various macroion charges $Q_\m{M}$ by solving 
the bulk Euler-Lagrange equations \Ref{eq:eleb}.

The macroion and the salt number densities are given by $\rho_\m{M} := \rho_\m{M_x}+\rho_\m{M_y}+\rho_\m{M_z}$ 
and  $\rho_\m{S} := \rho_\m{A}$, respectively. In order to detect the formation of liquid crystalline phases 
of the macroions, the equilibrium nematic order parameter for the director oriented relative to the $z$-direction,
\begin{equation}
   s_\m{M} := \frac{3}{2}\frac{\rho_\m{M_z}}{\rho_\m{M}}-\frac{1}{2} \in \left[-\frac{1}{2},1\right],
   \label{eq:sm}
\end{equation}
and the equilibrium biaxial order parameter,
\begin{equation}
   q_\m{M} := \frac{\rho_\m{M_x}-\rho_\m{M_y}}{\rho_\m{M}},
   \label{eq:qm}
\end{equation}
have been determined. The definition of $s_\m{M}$ agrees with the well-known scalar liquid-crystal 
order parameter $S = \langle P_2(\cos\theta)\rangle = \frac{3}{2}\langle(\cos\theta)^2\rangle-\frac{1}{2}$ 
because within the Zwanzig model only macroion orientations $\m{M_z}$ parallel ($\cos\theta = 1$) 
and $\m{M_{x,y}}$ perpendicular ($\cos\theta = 0$) to the $z$-axis are possible. 
$s_\m{M}$ vanishes in an isotropic phase ($\rho_\m{M_x} = \rho_\m{M_y} = \rho_\m{M_z}$),
whereas it is positive in a nematic phase with director parallel to the $z$-axis 
($\rho_\m{M_z} > \rho_\m{M_x},\rho_\m{M_y}$). A discrimination of the orientation $\m{M_z}$ leads 
to negative values of $s_\m{M}$.

It turned out that the biaxial order parameter $q_\m{M}$ vanishes throughout the whole inspected 
range of reduced chemical potentials $\mu_\comM^*$ and $\mu_\comS^*$, whereas the nematic order
parameter $s_\m{M}$ indicates either an isotropic fluid ($s_\m{M}=0$) or a nematic fluid 
($s_\m{M}>0$).

\begin{figure*}[thb]
   \includegraphics{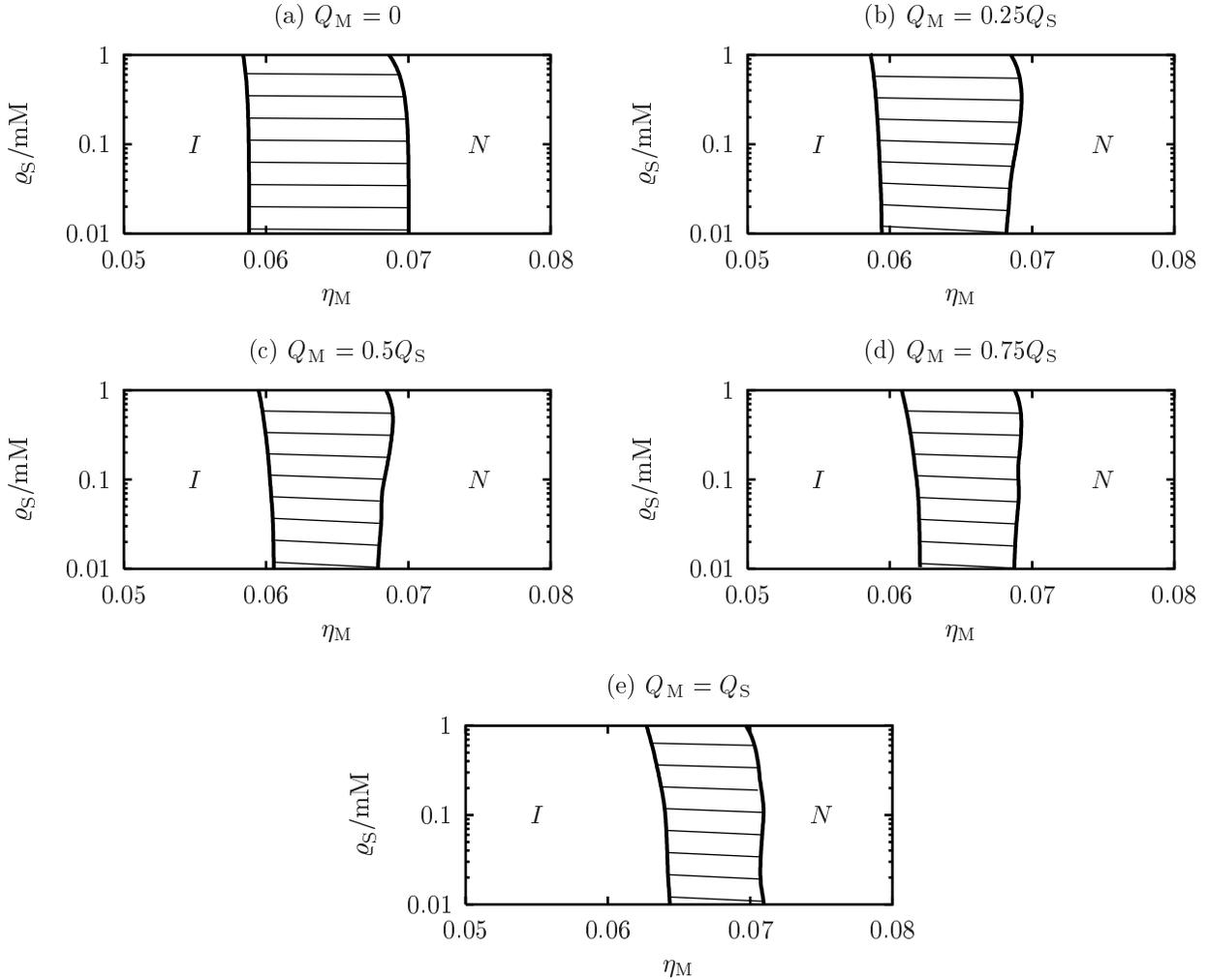}
   \caption{\label{fig:bpd}Bulk phase diagrams of mixtures of plate-like macroions 
           ($D_\m{M} = 20\,\ell_B$, $L_\m{M} = \ell_B$) and monovalent salt ($D_\m{S} = \ell_B$,
           $Q_\m{S} = -e$) for macroion charges $Q_\m{M} = 0$ (a), $Q_\m{M} = 0.25Q_\m{S}$ (b), 
           $Q_\m{M} = 0.5Q_\m{S}$ (c), $Q_\m{M} = 0.75Q_\m{S}$ (d), and $Q_\m{M} = Q_\m{S}$ (e) in 
           terms of the platelet packing fraction $\eta_\m{M} = \rho_\m{M}D_\m{M}^2L_\m{M}^{ }$ 
           and the salt density $\rho_\m{S}$ \cite{Units}. Note that $\rho_\m{S} = 1\,\m{mM}$ for 
           $D_\m{S} = 0.72\,\m{nm}$ corresponds to a salt packing fraction 
           $\rho_\m{S}D_\m{S}^3 = 2.2 \cdot 10^{-4}$. Coexisting states are connected by 
           (non-horizontal) tie lines. The model exhibits one isotropic ($I$) and one nematic ($N$) 
           phase separated by first-order phase transitions. The salt density of coexisting phases 
           is higher in the isotropic and lower in the nematic phase (Donnan effect) as can be inferred 
           from the negative slope of the tie lines. For increasing macroion charge, the isotropic and 
           the nematic binodals are shifted to larger macroion packing fractions $\eta_\m{M}$. With 
           increasing salt density $\rho_\m{S}$, the isotropic-nematic binodals for systems of charged 
           macroions bend towards lower values of the macroion packing fraction.}
\end{figure*}

Figure \ref{fig:bpd} 
shows phase diagrams for the parameters \cite{Units} (compare Fig. \ref{fig:geo}) 
$D_\m{M} = 20\,\ell_B \approx 14\,\m{nm}$, $L_\m{M} = \ell_B \approx 0.72\,\m{nm}$, 
$D_\m{S} = \ell_B \approx 0.72\,\m{nm}$, $Q_\m{S} = -e$ with $Q_\m{M} = 0$, $Q_\m{M} = 0.25Q_\m{S}$, 
$Q_\m{M} = 0.5Q_\m{S}$, $Q_\m{M} = 0.75Q_\m{S}$, and $Q_\m{M} = Q_\m{S}$ in terms of the macroion packing 
fraction $\eta_\m{M} = \rho_\m{M}D_\m{M}^2L_\m{M}^{ }$ and the salt density $\rho_\m{S}$.

\begin{figure}[thb]
   \includegraphics{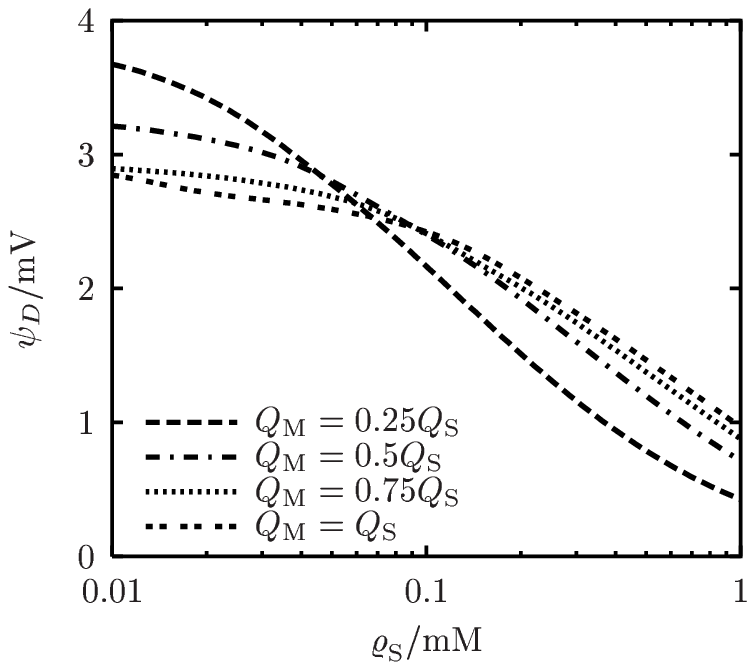}
   \caption{\label{fig:dp}Donnan potential $\psi_D$ (see Eq. \Ref{eq:psiD}) between the coexisting 
           nematic ($N$) and isotropic ($I$) bulk phases in mixtures of platelike macroions and salt 
           (see Fig. \ref{fig:bpd}). For fixed macroion charge $Q_\m{M}$, the Donnan potential 
           decreases with increasing salt density $\rho_\m{S}$. For salt densities 
           $\rho_\m{S} \approx 0.01\,\m{mM}$, $\psi_D$ decreases with increasing $|Q_\m{M}|$, whereas 
           for salt densities $\rho_\m{S} \approx 1\,\m{mM}$, $\psi_D$ increases with increasing $|Q_\m{M}|$.}
\end{figure}

One isotropic phase ($I$) and one nematic phase ($N$) are found separated by first-order phase transitions.
Whereas for coexisting phases $\eta_\m{M}$ is always smaller in the isotropic phase than in the nematic
phase, $\rho_\m{S}$ of coexisting phases is higher in the isotropic and lower in the nematic phase.
A similar displacement of salt from regions of large concentrations of charged macroions is known
as Donnan effect \cite{Adam73}. Whereas the original Donnan effect has been discovered in systems 
subdivided by membranes which are impermeable for macroions, here the density difference of the
macroions occurs due to two coexisting bulk phases. As for the case of membrane equilibrium, here
a Donnan potential $\psi_D$ (Eq. \Ref{eq:psiD}) maintains the density gradients between the coexisting
phases. Figure \ref{fig:dp}
shows its dependence on the macroion charge $Q_\m{M}$ and the salt density $\rho_\m{S}$. $\psi_D$
decreases with increasing salt density $\rho_\m{S}$. This tendency is intuitively expected as the 
Donnan effect becomes more pronounced with increasing macroion charge whereas increasing the salt 
density gives rise to a stronger screening of the macroion charge. For fixed salt density $\rho_\m{S}$ well
below $\approx 0.1\,\m{mM}$, $\psi_D$ decreases with increasing $|Q_\m{M}|$, whereas this behavior is reversed 
for fixed salt density $\rho_\m{S}$ above $\approx 0.1\,\m{mM}$.

Upon an increase of the macroion charge $|Q_\m{M}|$, the isotropic and the nematic binodals are shifted 
to larger values of the macroion packing fraction $\eta_\m{M}$. 
This may be qualitatively understood by introducing the notion of an effective shape, which, in the 
present case, for macroions is given by a hard core surrounded by a soft $\snorm{\cdot}$-sphere, i.e., 
a cube, with its linear extension proportional to $Q_\m{M}^2$ due to the pairwise Coulomb repulsion. 
For small macroion charges, the effective shape is still platelike whereas for highly charged colloids, 
the effective shape tends towards a cube leading to a shift of the two-phase region to larger macroion
packing fractions.

For fixed macroion charge as well as particle shape and increasing salt density $\rho_\m{S}$, the 
isotropic-nematic binodals in Fig. \ref{fig:bpd} bend towards smaller macroion packing fractions.
This behavior is expected intuitively, because high ionic strength causes strong screening which in 
turn leads to effectively quasi-hard platelets (see Fig. \ref{fig:bpd}(a)).

\vfill


\section{\label{sec:interface}Free interfaces}

\begin{figure}[thb]
   \includegraphics{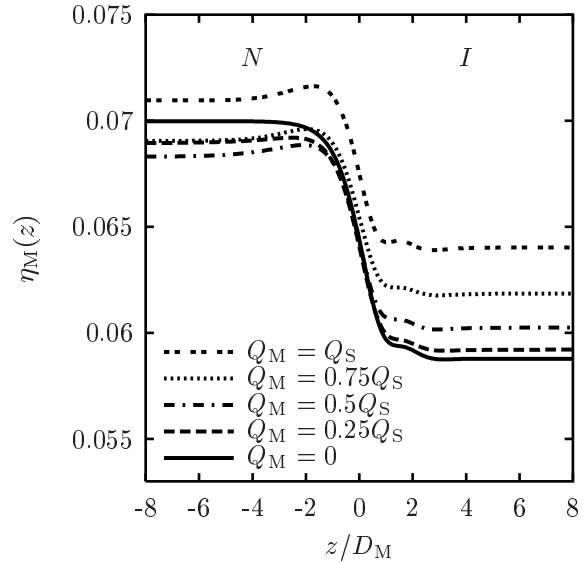}
   \caption{\label{fig:densities}Macroion density profiles at the free interface between 
           coexisting isotropic ($I$) and nematic ($N$) phases for the mixtures of platelike macroions and 
           salt as studied in Fig. \ref{fig:bpd} at nematic bulk salt density
           $\rho_\m{S}^{(N)} = 2.2 \cdot 10^{-5}\ \widehat{=}\ 0.1\,\m{mM}$ \cite{Units}. The interface position $z=0$ is 
           chosen such that $\rho_\m{M}(0) = \frac{1}{2}\left(\rho^\m{(I)}_\m{M} + \rho^\m{(N)}_\m{M}\right)$. 
           The interface width $\zeta$, inferred from the slope of the density profiles at the interface
           position $z=0$ (see main text), the Debye length $\kappa^{-1}$, and the bulk correlation lengths $\xi$,
           inferred from the exponential decay of the density profiles, decrease monotonically with increasing 
           macroion charge $|Q_\m{M}|$. The same behavior is found for the interfacial tension $\gamma$.
           See also Fig. \ref{fig:opp}.} 
\end{figure}

\begin{figure}[thb]
   \includegraphics{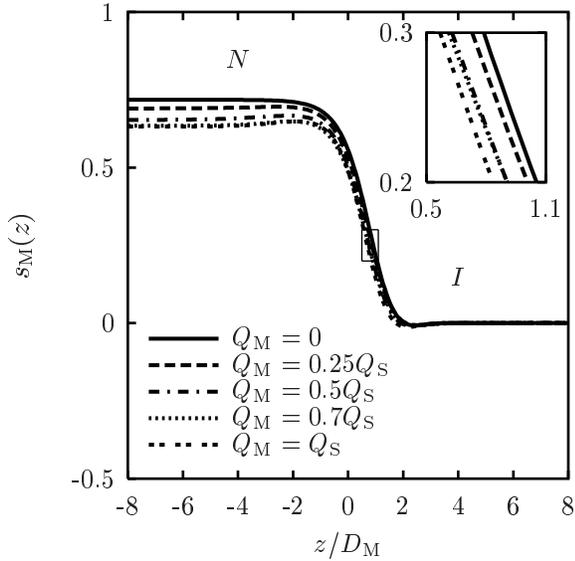}
   \caption{\label{fig:opp}Macroion nematic order parameter profiles $s_\m{M}$ (Eq. \Ref{eq:sm})
           at the free interface between coexisting isotropic ($I$) and nematic ($N$) phases for mixtures of
           platelike macroions and salt as shown in Fig. \ref{fig:bpd} at nematic bulk salt 
           density $\rho_\m{S}^{(N)} = 2.2 \cdot 10^{-5}\ \widehat{=}\ 0.1\,\m{mM}$ \cite{Units}. The position $z=0$ is 
           fixed by the choice $\rho_\m{M}(0) = \frac{1}{2}\left(\rho^\m{(I)}_\m{M} + \rho^\m{(N)}_\m{M}\right)$ (see
           Fig. \ref{fig:densities}). On the nematic side of the free interface, most of the macroions 
           lie parallel to the interface. For $z < 0$, $s_\m{M}(z)$ decreases with increasing macroion charge $|Q_\m{M}|$.
           The inset shows a detailed view of the steepest portions of the nematic order parameter profiles in the 
           range $s_\m{M}(z) \in [0.2,0.3]$ indicated by the frame.} 
\end{figure}

Based on the bulk properties provided in the previous section, we are now able to calculate the
density profiles at the free interfaces between the coexisting isotropic and nematic phases
by solving the spatially varying Euler-Lagrange equations (Eq. \Ref{eq:elefi}).
The density and order parameter profiles corresponding to the parameters used in the previous
section (see also Figs. \ref{fig:bpd} and \ref{fig:dp}) and to nematic bulk salt density 
$\rho_\m{S}^{(N)} = 2.2 \cdot 10^{-5}\ \widehat{=}\ 0.1\,\m{mM}$ \cite{Units} are depicted in Figs. \ref{fig:densities} 
and \ref{fig:opp}, 
respectively. The interface position $z=0$ is chosen such that $\rho_\m{M}(0) = 
\frac{1}{2}\left(\rho^\m{(I)}_\m{M} + \rho^\m{(N)}_\m{M}\right)$.

First, and most important, the formalism described in Sec. \ref{sec:formalism} renders stable 
free interfaces between coexisting bulk phases. This should be regarded as an accomplishment which
can be traced back to using the spatially varying screening factor $\kappa$ introduced in Eq. 
\Ref{eq:kappa}; trials with spatially constant $\kappa$ were not successful.

For a given density profile $\rho_\m{M}(z)$ (Fig. \ref{fig:densities}), the corresponding interface width 
$\zeta$ is defined as the spatial distance between the loci, where the tangent at the density profile
at position $z=0$ reaches the values of the nematic bulk density $\rho^\m{(N)}_\m{M}$ and the isotropic bulk
density $\rho^\m{(I)}_\m{M}$, respectively. This interface width $\zeta$ decreases monotonically with increasing 
macroion charge from $\zeta = 1.8D_\m{M}$ for $Q_\m{M} = 0.25Q_\m{S}$ to $\zeta = 1.3D_\m{M}$ for 
$Q_\m{M} = Q_\m{S}$ (see Fig. \ref{fig:densities}). The Debye length $\kappa^{-1}$ decreases monotonically from 
$\kappa^{-1} = 1.5D_\m{M}$ for $Q_\m{M} = 0.25Q_\m{S}$ to $\kappa^{-1} = D_\m{M}$ for $Q_\m{M} = Q_\m{S}$. Finally,
the bulk correlation lengths $\xi$ of the coexisting isotropic and nematic bulk phases, inferred from the 
exponential decay lengths of $\rho_\m{M}(z)-\rho^\m{(I,N)}_\m{M}$, also decrease monotonically upon increasing 
$|Q_\m{M}|$ and the values are by and large equal to those of $\kappa^{-1}$.

The nematic order parameter profiles $s_\m{M}(z)$ (Fig. \ref{fig:opp}) interpolate almost monotonically between
$s_\m{M}(\infty) > 0$ in the nematic bulk phase ($N$) and $s_\m{M}(\infty) = 0$ in the isotropic bulk phase 
($I$). Note that $s_\m{M}$ has been defined for a director in $z$-direction, i.e., platelets on the nematic
side ($z < 0$) are preferably oriented parallel to the free interface. At a fixed position on the nematic side 
($z < 0$), $s_\m{M}(z)$ decreases with increasing macroion charge $|Q_\m{M}|$. This behavior is consistent with 
the picture of an increasingly isotropic effective shape, introduced in the previous section. 

\begin{figure}[thb]
   \includegraphics{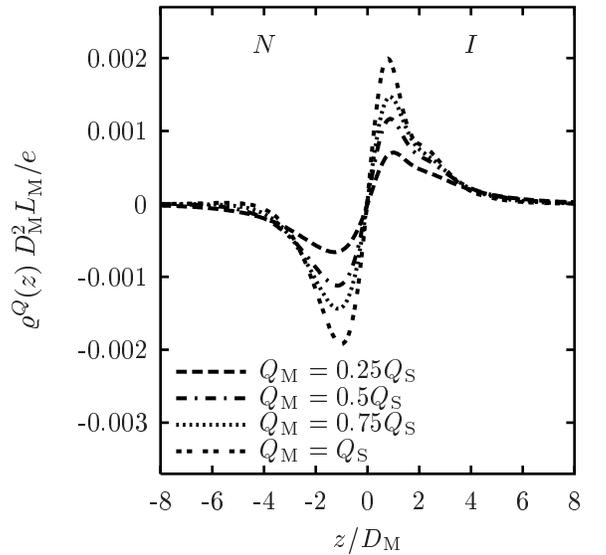}
   \caption{\label{fig:charge}Local charge density profiles (Eq. \Ref{eq:rhoc}) at the free 
           isotropic-nematic interfaces shown in Figs. \ref{fig:densities} and \ref{fig:opp} using 
           the same parameters and line code. Whereas \emph{global} charge neutrality holds, deviations 
           from \emph{local} charge neutrality occur near the interface with a negative charge density on 
           the nematic side ($N$) and a positive charge density on the isotropic side ($I$). The corresponding
           electrostatic potential profile is shown in Fig. \ref{fig:potential}.}
\end{figure}

\begin{figure}[thb]
   \includegraphics{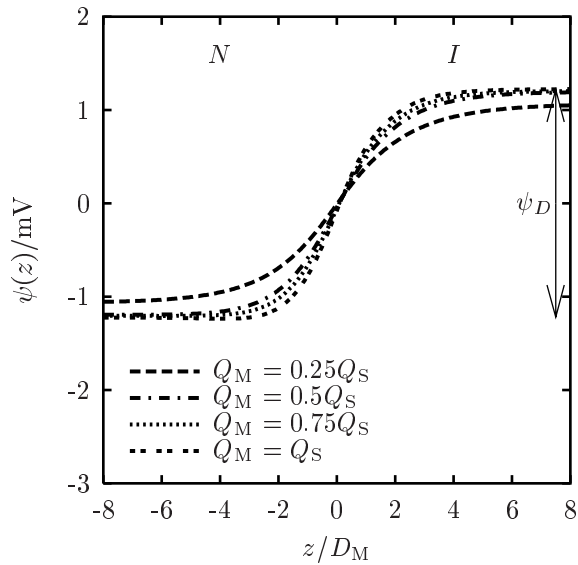}
   \caption{\label{fig:potential}Electrostatic potential profiles at the free isotropic-nematic 
           interfaces shown in Figs. \ref{fig:densities}-\ref{fig:charge}. The potential difference
           between the nematic and the isotropic bulk is given by the Donnan potential $\psi_D$ 
           displayed in Fig. \ref{fig:dp}; it is indicated for $Q_\m{M} = Q_\m{S}$.} 
\end{figure}

The charge density profiles $\rho^Q(z)$ (Eq. \Ref{eq:rhoc}) displayed in Fig. \ref{fig:charge}
show deviations from local charge neutrality within the interfacial region $-4D_\m{M} \lesssim z \lesssim 4D_\m{M}$.
A negative charge density occurs on the nematic side ($N$) and a positive charge density on the 
isotropic side ($I$). Such a local charging is necessary for the appearance of the non-vanishing Donnan 
potentials $\psi_D$ shown in Fig. \ref{fig:dp}. The full electrostatic potential profiles 
$\psi(z)$ are depicted in Fig. \ref{fig:potential}.
They increase monotonically from the macroion-rich nematic phase $N$ to the macroion-poor 
isotropic phase $I$, maintaining the density gradients occurring in the interface region. The
potential difference $\psi(\infty)-\psi(-\infty)$ equals the Donnan potential 
$\psi_D$ (see Eq. \Ref{eq:psiD} and Fig. \ref{fig:dp}).

The interfacial tensions $\gamma$ of the interfaces shown in Fig. \ref{fig:densities} decrease monotonically
from $\gamma = 1.33 \cdot 10^{-5}\ \widehat{=}\ 108\,\m{nN\cdot m^{-1}}$ for $Q_\m{M} = 0.25Q_\m{S}$ to 
$\gamma = 6 \cdot 10^{-7}\ \widehat{=}\ 5\,\m{nN\cdot m^{-1}}$ for $Q_\m{M} = Q_\m{S}$ \cite{Units}, which are
comparable to experimental findings for laponite suspensions \cite{Kooi01}. The corresponding
wetting parameters $\omega = (4\pi\gamma\xi^2)^{-1}$ \cite{Schi90} are in the range $6 \dots 340$.
If these values for $\gamma$ are indeed so small, the free isotropic-nematic interfaces are expected to be
strongly affected by capillary wavelike fluctuations which are not captured by the present theory. 

Calculating bulk phase diagrams and density profiles for macroion charges $Q_\m{M} \geq 1.25Q_\m{M}$
along the lines described above lead to unphysical results like, e.g., negative interfacial
tensions. The reason for this phenomenon is that the bulk phases determined in Sec. \ref{sec:bulk} are
assumed to be spatially homogeneous, whereas it can be shown that the equilibrium bulk states are spatially 
inhomogeneous for sufficiently large macroion charges (see Subsec. \ref{sec:Fex}).


\section{\label{sec:discussion}Discussion and Summary}

The numerical results for the bulk systems and the free interfaces presented in Secs. 
\ref{sec:bulk} and \ref{sec:interface}, respectively, are in good qualitative agreement 
with intuitive expectations. Within the density functional theory described in Sec. 
\ref{sec:formalism} one gains access not only to interfacial density profiles at free interfaces
between coexisting bulk phases but also to local charge densities and electrostatic
potential profiles.

Since here we have been interested in spatially homogeneous (isotropic or nematic) bulk phases, only very 
small macroion charges ($|Q_\m{M}| \leq e$) have been considered. Determining phase coexistence for
larger platelet charges is computationally more demanding because spatially inhomogeneous bulk 
phases are involved. Note that, according to Subsec. \ref{sec:Fex}, charges have to be interpreted as
effective charges.

A further difficulty related to the choice of the pair distribution functions $g_{ij}^{(\eta)}$ 
used in Eq. \Ref{eq:Fex1} may appear if, as for Eq. \Ref{eq:pdf}, the effective shape of
the macroions becomes more and more isotropic upon increasing the macroion charge: In this
case, the two-phase region between isotropic and anisotropic phase is shifted to
unrealistically large packing fractions. Therefore, the pair distribution functions must be chosen 
properly in order to yield \emph{anisotropic} effective macroion shapes up to large macroion charges.
Unfortunately, deriving \emph{analytical} expressions for pair distribution functions of platelike 
particles with inhomogeneous charge distributions is still a big challenge.

In conclusion, the density-functional theory of charged platelike particles developed
here can be regarded as a first step to understand qualitatively free interfaces between 
isotropic and nematic bulk phases of suspensions of charged platelets and salt. 
Here, we have constructed a density functional for charged platelike particles and 
applied it to a ternary mixture of platelike macroions and salt ions (Fig. \ref{fig:geo}) in 
the bulk and at free interfaces between coexisting isotropic and nematic phases. For sufficiently
small macroion charges, the bulk phase diagrams in terms of densities (Fig. \ref{fig:bpd}) exhibit 
one isotropic phase and one nematic phase. For increasing macroion charge and fixed salt densities, 
the two-phase coexistence region is shifted to larger macroion packing fractions. For fixed macroion 
charge and increasing salt density, the limit of quasi-hard platelets is approached. The 
Donnan potential between coexisting phases (Fig. \ref{fig:dp}) can be expressed in terms of
the particle chemical potentials gained from bulk structure calculations. Density and nematic 
order parameter profiles at free interfaces between isotropic and nematic phases at coexistence 
(Figs. \ref{fig:densities} and \ref{fig:opp}) show non-monotonic behavior. The value of the
nematic order parameter in the nematic bulk phase decreases upon increasing the macroion charge.
The width of the interface and the bulk correlation lengths are approximately given by the Debye 
length. The interfacial tension decreases upon increasing the macroion charges. Electrically charged 
layers form at the free interface (Fig. \ref{fig:charge}). The corresponding electrostatic 
potential profiles (Fig. \ref{fig:potential}) exhibit monotonic behavior. Investigations of spatially
inhomogeneous bulk phases are necessary in order to apply the theory to larger macroion
charges. Improvements of the present theory call for more accurate analytical 
expressions for the pair distribution function between charged platelike particles.






\begin{thebibliography}{00}
\bibitem{Units}   Dimensionless quantities are used in units of $\beta^{-1}=k_BT$ (thermal
                     energy), $e$ (proton charge), and $\ell_B=\frac{\beta e^2}{4\pi\epsilon_0\epsilon}$
                     (Bjerrum length) with the solvent dielectric constant $\epsilon$. For 
                     $T=300\,\m{K}$ and $\epsilon=78$ (water) the latter is $\ell_B \approx 
                     0.72\,\m{nm}$. In order to avoid a clumsy notation, the corresponding dimensional
                     quantities are denoted by the same symbol and can be recognized by the presence 
                     of units. Dimensional number densities are sometimes given in units of 
                     $1\,\m{mM} = 1\,\m{mol\cdot m^{-3}} \approx 6 \cdot 10^{23}\,\m{m^{-3}}$.
\end{thebibliography}
\end{document}